%% file: main.tex
\documentclass[10pt,journal,compsoc]{IEEEtran}

\ifCLASSOPTIONcompsoc
  \usepackage[nocompress]{cite}
\else
  \usepackage{cite}
\fi

\usepackage[many]{tcolorbox}
\newtcolorbox{myboxi}[1][]{
  breakable,
  title=#1,
  colback=white,
  colbacktitle=white,
  coltitle=black,
  fonttitle=\bfseries,
  bottomrule=0pt,
  toprule=0pt,
  leftrule=3pt,
  rightrule=3pt,
  titlerule=0pt,
  arc=0pt,
  outer arc=0pt,
  colframe=black,
}

\usepackage{ifthenx}
\newboolean{showcomments}
\setboolean{showcomments}{true} 
\ifthenelse{\boolean{showcomments}}
    {
         \newcommand{\neil}[1]{\textcolor{red}{{\it [Neil says: #1]}}}
                  \newcommand{\zane}[1]{\textcolor{green}{{\it [Zane says: #1]}}}
    }
    {  \newcommand{\neil}[1]{}
                \newcommand{\zane}[1]{}
    }

\newcommand{\corp}{DataCorp}

\ifCLASSINFOpdf
\else
\fi

\usepackage{todonotes}
\presetkeys{todonotes}{inline}{}
\usepackage{url}
\usepackage{booktabs}
\usepackage{comment}
\usepackage{graphicx}
\usepackage{array}
\usepackage{multirow}
\usepackage{enumitem}

\usepackage{hyperref}
\newcommand{\q}[1]{\emph{``#1''}}
\newcommand{\STAB}[1]{\begin{tabular}{@{}c@{}}#1\end{tabular}}

\begin{document}

\title{GDPR Compliance in the Context of \\Continuous Integration}

\author{Ze Shi Li,
        Colin Werner, 
        Neil Ernst,
        and Daniela Damian 
\IEEEcompsocitemizethanks{
\IEEEcompsocthanksitem Z.S. Li, C. Werner, N. Ernst, and D. Damian are with the University of Victoria, Victoria, BC, Canada, V8P 5C2. Email: \{lize,colinwerner,nernst,danielad\}@uvic.ca}
\thanks{Manuscript received December 19, 2019}}

\markboth{IEEE Transactions on Software Engineering}%
{Li \MakeLowercase{\textit{et al.}}: GDPR Journal}
%

\input{01_abstract.tex}

\maketitle

\IEEEdisplaynontitleabstractindextext

%
\IEEEpeerreviewmaketitle

\input{02_introduction.tex}

\input{03_background.tex}
\input{04_methodology.tex}

\input{05_problem_characterization.tex}

\input{06_operationalized.tex}
\input{08_discussion.tex}

\input{09_related_work.tex}

\input{10_implications.tex}
\input{11_threats_to_validity.tex}

\input{12_conclusions.tex}

\ifCLASSOPTIONcompsoc
  \section*{Acknowledgments}
\else
  \section*{Acknowledgment}
\fi

The authors would like to thank the employees and collaborators at \corp{} for their time and effort in working with our research team.

\ifCLASSOPTIONcaptionsoff
  \newpage
\fi



\bibliographystyle{IEEEtran}
\bibliography{main}
%



%
\begin{IEEEbiography}[{\includegraphics[width=1in,height=1.25in,clip,keepaspectratio]{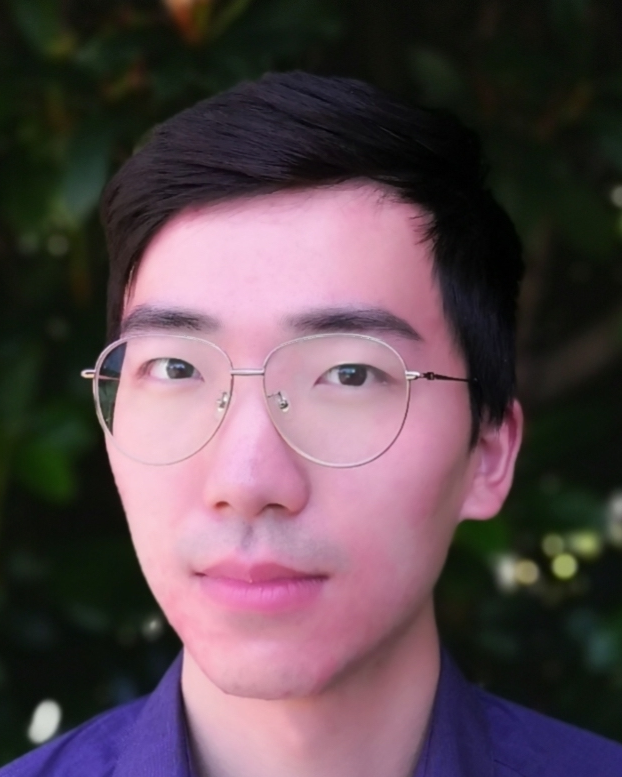}}]{Ze Shi Li}
is a master’s student at the University of Victoria. His research focuses on privacy, nonfunctional requirements, and continuous software engineering. His interests also include applying this research to industry.
\end{IEEEbiography}

\begin{IEEEbiography}[{\includegraphics[width=1in,height=1.25in,clip,keepaspectratio]{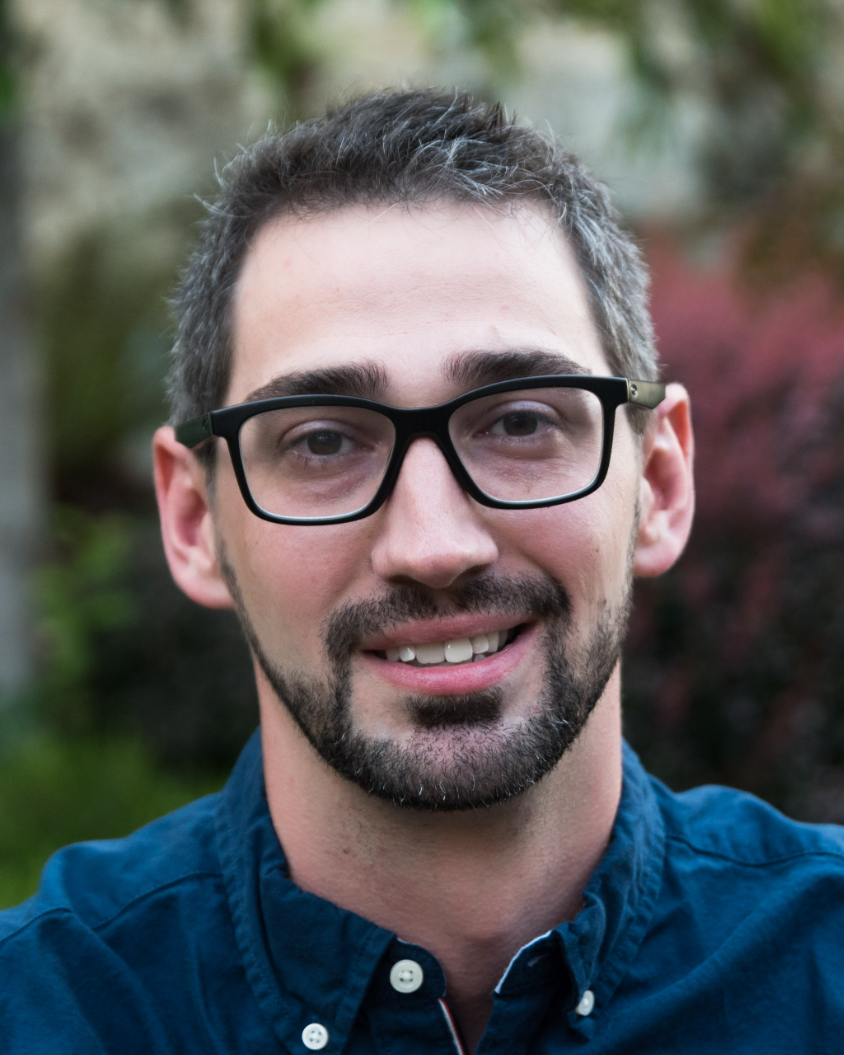}}]{Colin Werner}
is a Ph.D. student at the University of Victoria. His research interests revolve around solving practical industry-specific software engineering problems, with an emphasis on requirements engineering. He has extensive industrial experience working as a professional software developer prior to returning to school to pursue a Ph.D.
\end{IEEEbiography}

\begin{IEEEbiography}[{\includegraphics[width=1in,height=1.25in,clip,keepaspectratio]{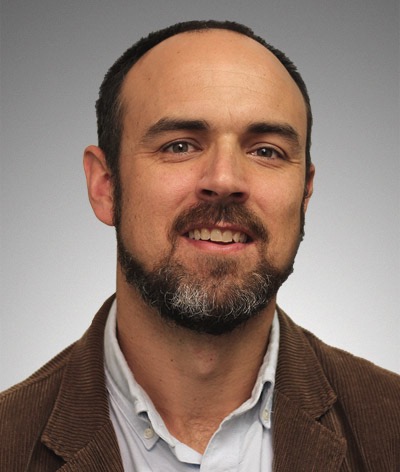}}]{Neil Ernst}
is a tenure-track assistant professor at the University of Victoria. He studies the intersection of software architecture and requirements. He leverages past experience consulting with large government stakeholders, and empirical datasets on software development and analysis. Current projects include software composition, technical debt in scientific software, and engineering data science systems. He has worked at the Software Engineering Institute at Carnegie Mellon University, the University of British Columbia, and received his PhD from the University of Toronto.
\end{IEEEbiography}

\begin{IEEEbiography}[{\includegraphics[width=1in,height=1.25in,clip,keepaspectratio]{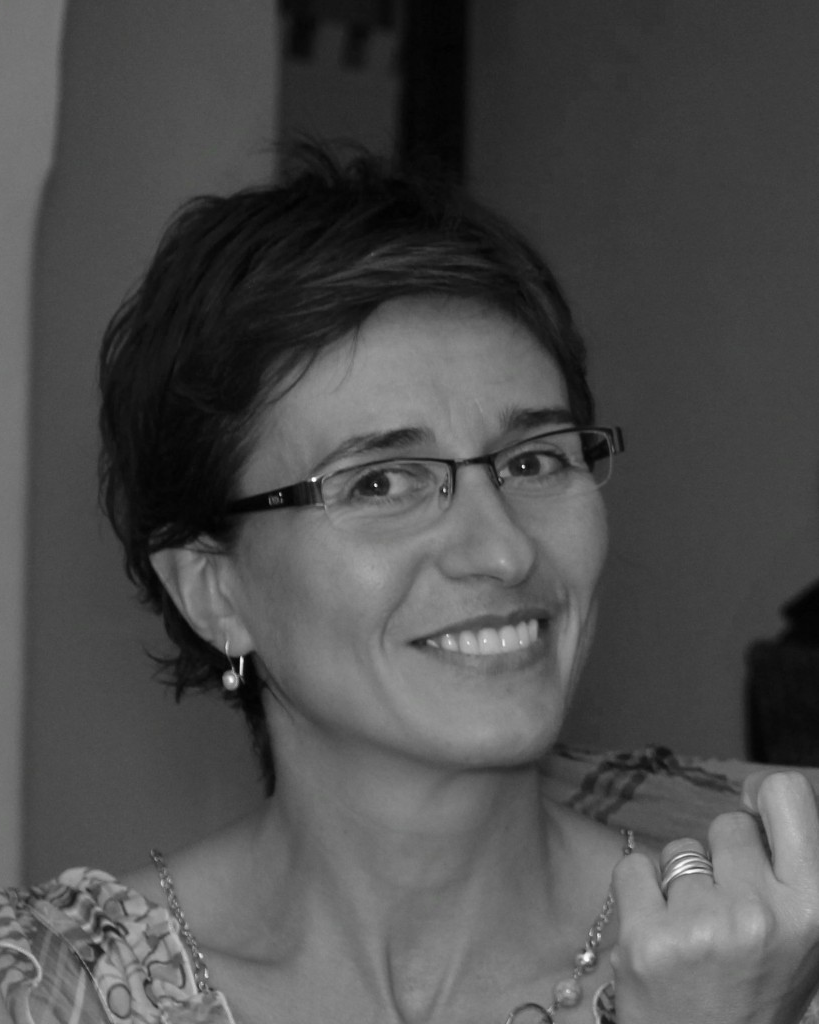}}]{Daniela Damian}
is a professor of software engineering in the University of Victoria’s Department of Computer Science, where she leads research in the Software Engineering Global interAction Laboratory. Her research interests include empirical software engineering, requirements engineering, and computer supported cooperative work. Her recent work has studied developers’ sociotechnical coordination in large, geographically distributed software projects, as well as stakeholder management in large software ecosystems. Damian received a Ph.D. from the University of Calgary. She has served on the program committee boards of several software engineering conferences, as well as on the editorial boards of IEEE Transactions on Software Engineering, Journal of Requirements Engineering, Journal of Empirical Software Engineering, and Journal of Software and Systems.
\end{IEEEbiography}

\end{document}

%% file: 01_abstract.tex
\IEEEtitleabstractindextext{%
\begin{abstract}
The enactment of the General Data Protection Regulation (GDPR) in 2018 forced any organization that collects and/or processes EU-based personal data to comply with stringent privacy regulations.
Software organizations have struggled to achieve GDPR compliance both before and after the GDPR deadline. 
While some studies have relied on surveys or interviews to find general implications of the GDPR, there is a lack of in-depth studies that investigate compliance practices and compliance challenges of software  organizations.
In particular, there is no information on small and medium enterprises (SMEs), which represent the majority of organizations in the EU, nor on organizations that practice continuous integration.
Using design science methodology, we conducted an in-depth study over the span of 20 months regarding GDPR compliance practices and challenges in collaboration with a small, startup organization.
We first identified our collaborator's business problems and then iteratively developed two artifacts to address those problems: a set of operationalized GDPR principles, and an automated GDPR tool that tests those GDPR-derived privacy requirements. This design science approach resulted in four implications for research and for practice.
For example, our research reveals that GDPR regulations can be partially operationalized and tested through automated means, which improves compliance practices, but more research is needed to create more efficient and effective means to disseminate and manage GDPR knowledge among software developers.
\end{abstract}

\begin{IEEEkeywords}
Privacy requirements, security, GDPR, continuous software engineering
\end{IEEEkeywords}}

%% file: 02_introduction.tex
\IEEEraisesectionheading{\section{Introduction}\label{sec:introduction}}
\IEEEPARstart{T}{he} General Data Protection Regulation (GDPR) \cite{gdpr_regulations} is a comprehensive EU privacy regulation that protects individual privacy and severely punishes privacy violations. 
The GDPR regulates any organization that is based in the EU \emph{or} collects and/or processes data from EU citizens. This broad scope implies a company from any other jurisdiction, such as the US or Canada, of any size, can easily fall within the purview of the GDPR. Apart from simply forbidding EU citizens from using their services \cite{npr_article, nbc_block}, these companies must adjust their software and information processing practices to comply with the new requirements imposed by the GDPR. 



A mistake with respect to requirements may be particularly costly for a small organization \cite{aranda_requirements_2007}.
A small organization, such as a startup, is likely to have fewer resources than a large organization to direct towards compliance and development, and may experience more difficulty with the GDPR.
A startup is typically in the stages of establishing a business model to consistently generate revenue \cite{ries_lean_2011}. 

To find a reliable source of revenue, a startup often moves quickly to find a suitable market, but may neglect non-functional requirements (NFRs) during the process \cite{gralha_evolution_2018}. Non-functional requirements focus on the quality of the startup's software, as opposed to specific functions \cite{glinz}.
As privacy legislation, the GDPR can be interpreted as an encompassing NFR, and GDPR regulations as a series of privacy NFRs.
When an organization neglects to prioritize NFRs early on, the organization is consequently prioritizing delivery of features over software quality \cite{gralha_evolution_2018}. 
%
However, as small organizations are often resource constrained and practicing continuous integration (CI) may be difficult \cite{ramesh_agile_2010}, important NFRs, such as privacy, may be neglected or less of a priority.

Many organizations are not yet GDPR compliant \cite{narendra_almost_2019,chantzos_gdpr_2019}, despite the regulation coming into full enforcement in 2018. 
To understand the reasons behind non-compliance, studies about the GDPR have often relied on surveys and interviews to study general compliance challenges \cite{sirur_are_2018, poritskiy_benefits_2019}. 
Unfortunately, there is not yet an established knowledge base of practical compliance challenges in literature.
In particular, no studies have either comprehensively explored privacy practices or challenges for small organizations that use continuous activities.

To investigate this gap, we conducted an extensive, 20 month study with a startup organization using design science (based on Hevner et al. \cite{hevner_design_2004}). 
Our partner organization, \corp{}, is a startup has a large number of users that reside in the EU. {\corp} makes extensive use of CI practices such as daily builds.
As 99\% of organizations in the EU are classified as SMEs \cite{eu_sme}, our study into a small organization is highly relevant for research and practice.

This paper makes these five contributions:
\begin{itemize}
    \item presents a detailed exploration on the practices and challenges of GDPR compliance in our collaborating organization; specifically, a mapping between context and compliance challenges is provided.
    \item presents a list of operationalized privacy requirements that are important to our collaborating organization and derived from three GDPR principles: integrity and confidentiality, data minimization, and storage limitation. 
    \item demonstrates how GDPR derived privacy requirements can be operationalized in an automated GDPR tool.
    \item provides empirical data from integrating a continuously running, automated GDPR tool to raise awareness about potential GDPR exposures, and obstacles of continuous compliance.
    \item lists four research implications and four practitioner implications based on our investigation.
\end{itemize}

\begin{figure*}[t!]
  \centering
    \includegraphics[width=18cm]{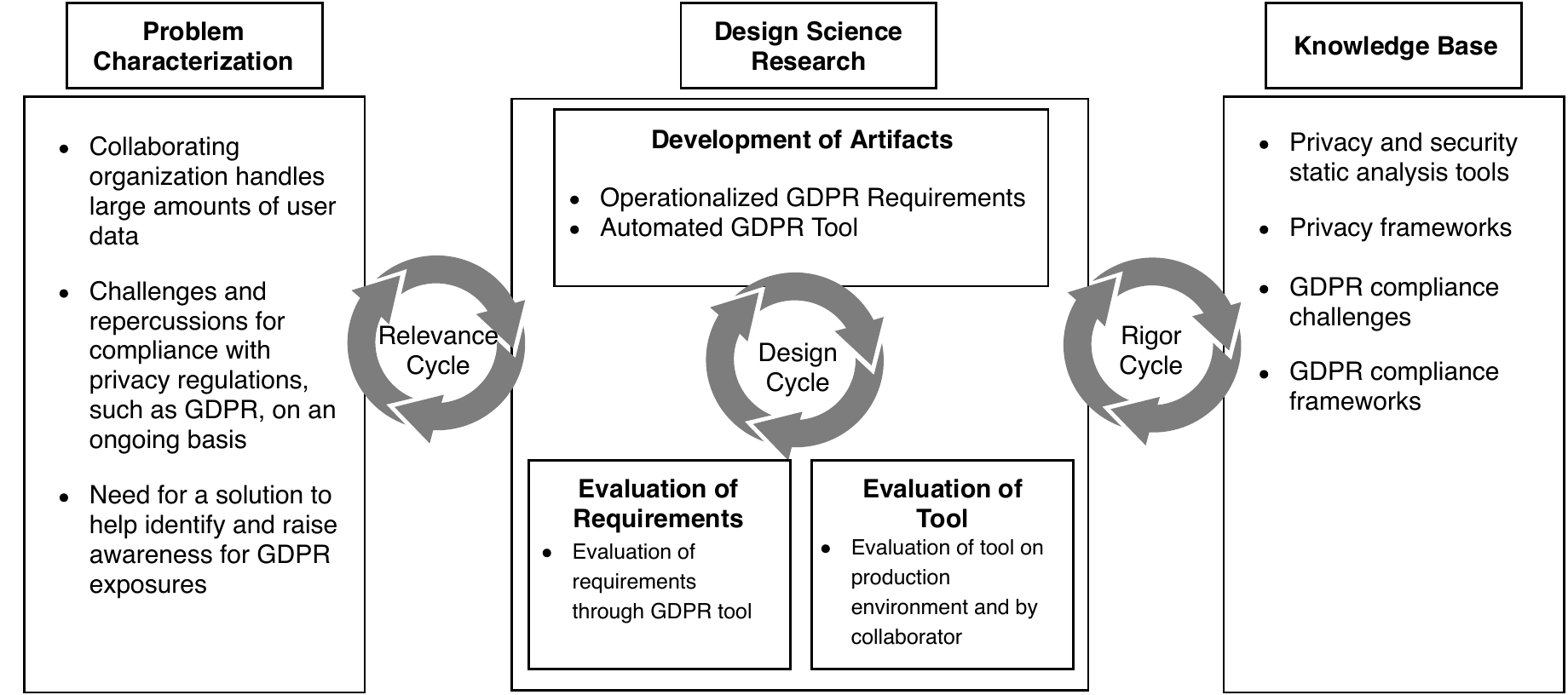}
  \caption{Design Science Methodology, based on Hevner et al. \cite{hevner_design_2004}. Design science is a research process  emphasizing characterizing relevant problems (left side), iteratively developing and evaluating artifacts that potentially solve those problems (middle), and ensuring the findings and artifacts are rigorous (right side).}
  \label{fig:design_science}
\end{figure*}

%% file: 04_methodology.tex
\section{Methodology}
We begin by outlining our research method, including specific processes followed, and the describe the findings of those processes in subsequent sections. 
\subsection{Design Science Methodology}
The two driving forces of our research were the gap in knowledge of compliance practices and challenges of small organizations practicing continuous activities and our collaborating organization's urgency to achieve GDPR compliance. 
With operations in the EU, our collaborator inherently expressed interest in researching effective means to become and remain GDPR compliant.

Through design science research \cite{sedlmair_design_2012, hevner_design_2004}, that involved a mixed-methods approach, involving ethnographic informed methods, including participant observation and interviews spanning over 20 months, we acquired first hand insight on compliance practices and challenges experienced by a startup organization and studied how an automated tool may help ensure an organization's compliance. 
We chose design science because it emphasizes the importance of finding relevant problems in the investigated organization and developing and evaluating artifacts to reduce the burden of these problems. 
Figure \ref{fig:design_science} depicts the elements of our design science methodology. 
In particular, the left part of Figure \ref{fig:design_science} depict the findings of our problem characterization, which serve to uncover important and relevant business problems \cite{hevner_design_2004}.
The design science artifacts produced in our research must be relevant for our collaborator and rigorously evaluated \cite{hevner_design_2004}.
The rigor cycle of our research was attained in two ways. One, internal validation, and two, external validation through an analysis of existing literature regarding the GDPR, and methodologies and tools designed to handle privacy. 
In particular, we ensured external validation by reviewing existing literature on GDPR implications for organizations, and privacy frameworks and methodologies designed to help achieve compliance. 

For internal validity, we note that the nature of the design science cycle --- from problem to solution and back to validation in practice --- ensures that the solution has relevance to (at least) our partner. However, we also ensured credibility and analyzability of the data by:
\begin{itemize}
    \item the primary researcher embedded with {\corp} maintained a researcher diary of observations, with entries of each observation day;
    \item we thematically coded all text artifacts, expanded the relevant challenges with our observational data and diary notes, and then validated these challenges for relevance with the primary company contact;
    \item we conducted iterative, ongoing member checking with three members of the company, as we developed our analysis;
    \item we member checked the final conclusions, described in this paper, with the primary contact at {\corp}.
\end{itemize}

\subsection{Research Setting}
Our collaborating organization, {\corp}\footnote{Real name and some identifying details have been changed for confidentiality.}, is a data gathering and analysis startup. Some data is gathered from EU users.
During our study, {\corp} experienced immense growth, starting from a handful of employees and ending up with several dozen.
Part of {\corp}'s business is collecting millions of data points every day, which include many users from the EU.
Since the GDPR prescribes GDPR compliance from any organization that collects personally identifiable data from any EU citizen, {\corp} maintained the obligation to become compliant by the GDPR deadline. 
Data are pseudo-anonymized when collected by {\corp}, as a precautionary measure to protect privacy.
For development, {\corp} uses CI tools, such as Jenkins, to automate software build and deploy software to production.
After code is committed and pushed to source control, {\corp}'s deployment pipeline builds the code and runs automated tests against the code, if pertinent tests exist.
{\corp} makes extensive use of cloud hosting solutions for data storage and analysis. 
As such, {\corp} has multiple categories of ``partners": 1) customers who receive data from {\corp}; 2) third-party services who provide infrastructure to collect, store, and process data; 3) partners who facilitate data collection.

As part of our research, the first author led a mixed-methods approach involving ethnographic informed methods, including participant observation and interviews, whereby he became part of the team and its activities. 
He spent one to two days per week in {\corp}'s offices.
To acquire a reasonable perspective of {\corp}'s work, the co-author participated in meetings, such as planning and retrospective meetings and performed tasks such as creating documentation for data flow. 
We also received access to some of {\corp}'s source control repository, project management tools, and infrastructure hosted in the cloud. 
Furthermore, we interacted with employees, conducting interviews, as well as learning and observing the organization's processes. 

These types of activities facilitated the increase in our awareness on how {\corp} planned work, developed code, tested software, and types of tools used to support {\corp}'s work.
In addition, analyzing project management tasks gave us insight into the type and distribution of tasks, as well as the amount of preparation conducted for GDPR compliance.
Ultimately, our study facilitated our grasp on the practices utilized by {\corp} for compliance and active challenges that hinder its compliance ability.

\subsubsection{Problem Characterization}
The problem characterization step of our research sought to understand the challenges experienced by {\corp}.
Hence, the first co-author spent one to two days per week at {\corp} throughout the study. 
During the initial eight months at {\corp}, the author participated in at least one meeting per month, conducted interviews with nine employees, observed numerous discussions, and conversed with essentially every {\corp} employee.
Based on our problem characterization, we identified relevant problems in the organization and found potential causes of these problems.



\subsection{Development and Evaluation of Artifacts}
To mitigate the difficulties found in our problem characterization, especially regarding awareness and time, our design science research methodology produced two iteratively developed and evaluated artifacts as shown in Figure \ref{fig:design_science}: privacy requirements and GDPR tool.  


\begin{table}[]
    \caption{Participant Role and Experience}
    \centering
    \begin{tabular}{c|c|c} \toprule
    Id  & Role & Time in Organization  \\ \midrule
    P1  &  Developer  & Less than 5 years \\
    P2  &  Developer & 5 or greater \\
    P3  &  Manager & 5 or greater \\
    P4  &  Manager & 5 or greater\\
    P5  &  Developer & 5 or greater \\
    P6  &  Developer & Less than 5 years \\
    P7  &  Developer & Less than 5 years \\
    P8  &  Developer & Less than 5 years\\
    P9  &  Manager & 5 or greater \\ \bottomrule
    \end{tabular}

    \label{tab:participant_info}
\end{table}

%% file: 05_problem_characterization.tex
\begin{table*}[t]
    \caption{Relationship between observed GDPR challenges and organizational context of {\corp}. Contextual factors (rows) \emph{contributes to} one or more specific GDPR challenges (columns). These challenges are discussed in \S \ref{problem_characterization_challenges}.}
    \centering
    \label{context_challenges}
    {\renewcommand{\arraystretch}{1.2}
    \begin{tabular}{lp{6.25cm}>{\centering\arraybackslash}p{0.95cm}>{\centering\arraybackslash}p{0.95cm}>{\centering\arraybackslash}p{0.95cm}>{\centering\arraybackslash}p{0.8cm}>{\centering\arraybackslash}p{0.95cm}>{\centering\arraybackslash}p{0.95cm}>{\centering\arraybackslash}p{1.2cm}}
    \toprule
    ~ & ~ & \multicolumn{7}{c}{\textbf{Challenges}} \\
    ~ & ~ & \multicolumn{3}{c}{\textbf{Awareness and Knowledge}} & \multicolumn{1}{c}{\textbf{Testing}} & \multicolumn{3}{c}{\textbf{Business and Workflow}} \\
    ~ & ~ & \STAB{\rotatebox[origin=l]{75}{\parbox[t][0.8cm][t]{2.2cm}{Understanding the GDPR}}} & \STAB{\rotatebox[origin=l]{75}{\parbox[t][1cm][t]{2.2cm}{Becoming aware of new regulations}}} & \STAB{\rotatebox[origin=l]{75}{Educating users}} & \STAB{\rotatebox[origin=l]{75}{\parbox[t][1cm][t]{2.2cm}{{Manual testing of privacy requirements}}}} & \STAB{\rotatebox[origin=l]{75}{\parbox[t][1cm][t]{2.2cm}{Checking for GDPR compliance}}} & \STAB{\rotatebox[origin=l]{75}{\parbox[t][0.8cm][t]{2.2cm}{Long term GDPR compliance}}} & 
    \STAB{\rotatebox[origin=l]{75}{\parbox[t][1.3cm][t]{2.2cm}{Ensuring compliance from customers and processors}}} \\
    \midrule
    \multirow{14}{*}{\textbf{Context}} & Number of GDPR Regulations & X & ~ & ~ & ~ & ~ & ~ & ~ \\ 
    ~ & Ambiguity of GDPR & X & X & ~ & ~ & ~ & ~ & ~\\
    ~ & Lack of legal training & X & ~ & ~ & ~ & ~ & ~ & ~\\
    ~ & Lack of privacy experience & X & ~ & ~ & ~ & ~ & ~ & ~ \\
    ~ & 
    Conflicting advice from experts
    & X & ~ & ~ & ~ & ~ & ~ & X  \\
    ~ & Nature of business & ~ & ~ & X & ~ & ~ & X & X \\
    ~ & Size of organization & ~ & X & ~ & X & X & X & X \\
    ~ & Lack of time  & ~ & ~ & ~ & X & X & X & ~  \\
    ~ & Increased growth of infrastructure and data & ~ & ~ & ~ & X & X & ~ & X  \\
    ~ & Data subject rights granted by the GDPR & ~ & ~ & ~ & X & ~ & ~ & X  \\
    ~ & Making existing systems compliant & ~ & ~ & ~ & ~ & X & X & X \\
    ~ & Lack of shared understanding & ~ & ~ & ~ & ~ & X & X & ~  \\
    \bottomrule
    \end{tabular}
    }
\end{table*}

\section{Problem Characterization} \label{problem_characterization_challenges}
Our design science research first establishes the relevance of our research to an actual business setting at {\corp}. 
As part of problem characterization, we interviewed nine employees, which consisted of developers and managers.
Table \ref{tab:participant_info} lists each interviewee's primary role and time spent in {\corp}.
Due to our ethics guidelines and NDA signed with {\corp}, we anonymized each interviewee.
In contrast, a manager represents someone whose primary focus is managing developers or other employees.
A developer represents someone who mostly works in development, testing, or operations. 
A ``manager" may still perform development tasks as {\corp} is a startup and employees often have multiple responsibilities. 
The interview questions from the appendix section lists the template of questions that we asked each interviewee.
Since we conducted observations and interviews, we could corroborate our findings to define the problem instance.   
During the interviews, we also ran a survey whereby each interviewee was asked to prioritize NFRs based on a list of thirteen NFRs. 
The survey entailed two iterations.
The first iteration involved ranking each NFR based on an interviewee's role, whereas the second was from the perspective of the business.

We identified three main challenges at {\corp} that hinder GDPR compliance:
\begin{enumerate}
	\item reliance on manual GDPR tests,
	\item limited awareness and knowledge of privacy requirements, and
	\item balancing GDPR compliance in a competitive data business.
\end{enumerate}

Table \ref{context_challenges} maps our observed context and circumstances to the challenges. 
We describe the challenges in more detail in the following subsections.


\subsection{Reliance on Manual GDPR Tests} \label{manual_testing_challenge}

{\corp} heavily relies on manual tests to conduct GDPR compliance testing, leading to significant time pressure for an individual to verify compliance. 
Furthermore, continuous growth of {\corp}'s infrastructure as a result of maturing further intensified the challenge with manual testing given low allocatable time. 

When our research began, {\corp} was much smaller in size (i.e. a handful of employees as opposed to several dozen) and {\corp's} employees often had a multitude of responsibilities and were stressed by time constraints.
We often heard employees say ``I would...but I have no time" (P2) or ``I wish I had more time" (P6).
{\corp} uses some automated tests,
such as ensuring its app store privacy policy matches the version in the internal company repository.
However, manual tests are the predominant strategy to test privacy requirements. 
If a privacy requirement stipulated a stoppage of data collection for a specific data parameter, a developer would need to manually check a database to verify the data parameter was no longer collected by the organization's system. 

However, manual tests are laborious, error prone, and time consuming \cite{Dustin:1999:AST:310674}. 
It is very easy for a developer to check the wrong database or run the wrong query.
Any erroneous manual test will ultimately result in rework or retesting of the privacy requirement, as well as hindering compliance.
To check that a privacy requirement still applies after every change to the database, a developer would have to conduct the same type of manual test after every code change. 

As {\corp} matured, its infrastructure and data also experienced immense growth. 
{\corp} cannot continue its manual approach to testing privacy requirements; either developers are redirected from other work or system elements are ``assumed" to be GDPR compliant.
In particular, {\corp} verifying the GDPR compliance of {\corp}'s infrastructure is particularly time-consuming.
Manually finding GDPR exposures has the potential benefit of a human interpreting a subjective scenario, but manual testing is slow. 

{\corp} relies on a multitude of third party services like Amazon Web Services (AWS), Google Cloud Platform (GCP), and Azure; the organization has many infrastructure resources hosted by these third party services. 
For instance, {\corp} hosts more than 50 databases and over one hundred servers on a single third-party cloud service.
It is arduous and tedious for a developer to manually review all those databases. 
Moreover, the quantity of resources also experienced rapid growth; the number of servers on one service increased 14\% over a 5 month period.
Hence, a developer tasked with uncovering a GDPR exposure in {\corp}'s entire infrastructure may require substantial time. 

Data subject rights granted by the GDPR also reduced the allocatable amount of time at {\corp}.
For example, a user may request an organization to provide all existing data about the user, 
the organization must terminate data collection and delete a user's data upon request even if the user gave prior consent to data collection.
Thus, soon after the GDPR deadline, {\corp} began receiving emails and requests from various users asking to stop collecting their data.
However, {\corp} has a manual termination process that requires an individual's response to each user. As explained by P9 \q{When a user send a request to opt out to us, the emails come to me and I have to tell them how to opt out}, the organization must respond to each individual user.
If the organization receives a plethora of requests per day, P9 would have to help satisfy each user, which may inhibit other important work given each employee's busy schedule. 


\subsection{Limited Awareness and Knowledge of Privacy Requirements}
It can be difficult for {\corp} to properly identify privacy problems, due to the complexity and magnitude of the GDPR and inexperience dealing with privacy regulations. 
Additionally, the lack of awareness of new privacy regulations may inhibit long term privacy compliance. 
{\corp} must also manage privacy awareness of users to collect data.

Ideally, each {\corp}'s employee is knowledgeable and reasonably understands the GDPR, but attaining a sufficient understanding is difficult.
The GDPR consists of ninety-nine articles and one hundred seventy-three recitals \cite{gdpr_regulations}, but the entire GDPR is written in legal speak.
For lawyers, the GDPR may be straightforward, but {\corp}'s employees are not well-versed in legal language nor have specific privacy training.
Hence, {\corp}'s developers are unsure about the requirements dictated by the GDPR, which may prevent effective treatment of a privacy NFR.
In addition, GDPR regulations are often ambiguous \cite{cool_impossible_2019}, which further hindered understanding.
For example, \q{[Evaluating GDPR compliance of tools] is difficult because I am not an expert in the GDPR. 
} (P1) and \q{Interpreting the rules and regulations [was challenging]. The rules weren't clear on what can be collected and what is considered private} (P9). 
The inexperience with privacy regulations also reduced the ability to share GDPR knowledge with each other.
Furthermore, some GDPR compliance guides lack details or contain inaccuracies, which could create misinterpretations or misunderstanding. 
{\corp} also indicated that some external consultants even provided contrasting answers to the same question.

For long term compliance, {\corp} should \q{Stay up to date with the regulations. Put efforts in research and implement the changes} (P7).
Yet, none of our 9 interviewees could definitively describe an upcoming privacy regulation, but a manager rightly speculated that the US would eventually pass privacy laws: \q{No [not aware of any new regulations], but US will probably adopt something similar to the GDPR} (P9).
Not knowing an upcoming privacy regulation does not have a direct negative affect on current GDPR work, but awareness of forthcoming privacy NFRs may prevent duplicate work and simplify future compliance adoption.
Even staying up to date with the GDPR may be difficult: \q{[A large challenge is knowing] changes to the GDPR. Especially minor changes [and amendments] can be difficult for companies to find out} (P4).

Another difficulty of managing awareness is educating users on {\corp}'s data collection purposes.
As explained by P9 \q{a user needs to be educated on why we are collecting data}.
Without sufficient explanation, a user may decline the terms of service or report the organization to a data protection agency. 
As users play a pivotal role to {\corp}'s business, {\corp} must sufficiently communicate and inform collection purpose and get consent from users.

Regarding privacy work, {\corp} had an unequal distribution of tasks as managers and a few specific developers seemed to receive the bulk of tasks.
Hence, many employees felt insignificant impact from the GDPR.
In our NFR survey, managers also felt privacy was significantly more important to {\corp}'s business than developers. 
Finally, CI may provide a couple advantages for {\corp} for achieving compliance, namely quick release and feedback: \q{Through CI, [redacted] can be generated, modified, and fixed within a couple of hours} (P5) and \q{[allows involvement] with external stakeholders} (P9).
However, these compliance benefits may be contingent on employees possessing a sufficient level of GDPR knowledge.
For a developer to implement fast changes and receive rapid feedback, the developer has to recognize the expectations of the GDPR.

\subsection{Balancing GDPR Compliance in a Competitive Data Business}
Due to {\corp}'s business approach, {\corp} is affected by the GDPR regulations.
Earning the trust of users and receiving consent is paramount to the success of the organization, but even if a user consents, \q{there may be a regulator who says we can't collect this data} (P9).
Hence, despite {\corp}'s best efforts to justify its data collection and takes adequate steps to safeguard its systems, a national privacy regulator could decide that {\corp} is not allowed to collect data.



Complicating matters for {\corp} is that it is a small organization with many competitors.
For instance, \q{Staying competitive in terms of [volume of data] we collect and present, while respecting privacy concerns of anonymization} (P5).
To stay competitive against other companies, {\corp} needs to continue increasing the amount of data collected from users.
Therefore, {\corp} needs to balance GDPR requirements and {\corp}'s business. 

Since {\corp}'s system already exists, becoming compliant means re-designing large aspects of the system to comply with the GDPR.
Aspects of {\corp}'s system has existed for years.
At this stage, it is challenging to modify elements that affect the system architecture.
P6 admitted that \q{building a GDPR compliant product is easier than making a legacy system GDPR compliant}.

Due to the GDPR's emphasis on shared responsibility between controllers and processors, {\corp} must also vet its partners: \q{[It's challenging] making sure that partners who receive data are compliant} (P3).
Furthermore, based on GDPR data erasure policies, if {\corp} receive a request to delete a user's data, {\corp} must also forward the request to \emph{every} partner who received the user's data and ensure that these partners also comply with the user's request. 
Essentially, {\corp}'s compliance is also tied to the compliance of {\corp}'s partners. 

In addition, a lack of shared understanding in {\corp} is another contributing factor to compliance difficulty. The lack of shared understanding often manifested itself in unsafe assumptions.
For example, we saw instances of developers \emph{assuming} elements are secure and compliant.
In addition, a developer explained that GDPR compliance was not a significant concern because their work dealt with data that was already pre-processed. 
The developer assumed that prior processes contained safeguards and checks that would ensure the data is GDPR compliant. 
However, the developer's assumption implies the organization has mechanisms in place to ensure this assumption is accurate and traceable, which \corp{} does only partially.

%% file: 06_operationalized.tex
\section{Design Science Artifacts } \label{operationalized_gdpr_requirements}

\begin{table*}[h]
\centering
\caption{Mapping of GDPR Principles to Privacy Requirements}
\label{tab:requirements}
{\renewcommand{\arraystretch}{1.1}
\begin{tabular}{lp{13cm}}
\toprule
{\textbf{GDPR Principle}}  & {\textbf{Privacy Requirement}} \\ 
\midrule
\multirow[t]{15}{*}{Integrity and Confidentiality}  & A database must be encrypted  \\ 

 & Each server must exist with a purpose  \\ 
 &  Each server without purpose must be removed \\ 
 &  Each server must have a corresponding cloud firewall \\ 
 & Each server storage must be encrypted  \\
 & Each server storage must exist for a purpose  \\
 & Each cloud firewall must use secure protocols inbound and outbound \\
 & Each cloud firewall must limit access to reliable sources  \\
 & Each cloud firewall must limit outbound communication to reliable sources \\
 &  Each load balancer must use end to end encryption  \\
 &  Each load balancer must use secure protocols  \\
 & Each cloud storage resource must be encrypted \\
 &  Each cloud storage resource must limit access to reliable sources \\
 & Each cloud storage resource must limit modification and deletion to reliable sources \\
 & Each access management resource must not grant unconditional permissions  \\
 & Each access management resource must not grant permissions to unconditional resources   \\
 & Each router must limit outbound communication to reliable sources \\ 
Data Minimization  & Each database must not collect personal data types outside an organization's data collection purpose \\ 
Storage Limitation  & Each database tuple must not live indefinitely  \\
\bottomrule
\end{tabular}}
\end{table*}

Based on the problem characterization step of our design science approach, {\corp} dealt with three main challenges --- reliance on manual testing, limited awareness, and GDPR compliance in a competitive environment.
We determined that reliance on manual GDPR tests is the most important and tractable challenge, especially as employees deal with constraints to time and long term manual testing of the GDPR is unsustainable.
Over reliance on manual testing was a bothersome challenge particularly pertinent to {\corp}'s growing infrastructure.
In contrast, the other two challenges are less likely to be directly solved by software-only solutions. 


We thus embarked on constructing artifacts (which include processes, tools, and models) as part of the middle phase of the design science cycle shown in Figure \ref{fig:design_science}. The goal is to construct artifacts to help reduce the problem of over-reliance on manual testing at {\corp}. Our approach is to automate the manual testing to reduce staff effort and make the process of compliance more repeatable and automatic. To do this, we first determine which GDPR principles are most amenable to automated testing, within the specific context of our partner. We analyzed {\corp}'s infrastructure and the GDPR principles and found three pertinent GDPR principles to {\corp}'s infrastructure. These are shown in Table \ref{tab:requirements}.

We operationalized these three GDPR principles into specific privacy NFRs that apply to {\corp}.
The privacy NFRs were then automated in a custom-built GDPR tool to raise awareness about potential GDPR exposures and continuously verify whether {\corp}'s infrastructure is satisfying these NFRs. 

Over a period of six months, we iteratively developed and evaluated our design science artifacts.
Since development of artifacts was heavily influenced by compliance challenges at {\corp}, it was paramount that {\corp} provided guidance and feedback in the evaluations of our artifacts. 
Ultimately two design science artifacts were produced: privacy requirements operationalized from GDPR principles and an automated GDPR tool.

\subsection{Operationalizing GDPR Principles into Privacy Requirements}
The GDPR has six main data processing principles 1) lawfulness, fairness and transparency 2) purpose limitation 3) data minimisation 4) accuracy 5) storage limitation 6) integrity and confidentiality \cite{gdpr_regulations}. 
Accountability is another primary GDPR principle, but accountability's purpose is requiring organizations to adhere to GDPR regulations and demonstrate compliance. 

Our first design science artifact is our list of privacy requirements operationalized from GDPR principles, shown by Table \ref{tab:requirements}. 
Based on input from {\corp} and our own observations, the integrity and confidentiality principle was the most important candidate to be operationalized (by operationalized, we mean the process of confirming whether the NFR is automatically satisfied). 
The purpose of this principle is to ensure that the organization is adequately handling personal data, and safeguarding that data from malicious attacks or accidental misappropriation. 
For example, one example of a specific requirement based on this GDPR principle is that databases and servers must be encrypted. This was explained to us by two different employees: \q{I added more encryption to the databases} (P6) and \q{[I worked on] disk and storage encryption} (P2).
Prior to our study, some employees were assigned related tasks, but there was no systematic strategy of verifying each infrastructure element, leading to potential privacy exposures in the system. 
Moreover, section \ref{manual_testing_challenge} elaborates on {\corp}'s extensive cloud-based infrastructure, that makes manual testing of every infrastructure resource an arduous process.  


Our second operationalized principle, storage limitation, as shown by Table \ref{tab:requirements}, represents the idea of keeping data no longer than necessary. 
An organization must ensure that it has a process to remove a datum after a period of time.
For example, a datum is automatically removed after a year.
In {\corp}'s situation, the multitude of data ideally is automatically removed after a specified time frame.


Similarly, the data minimization principle instills the notion that personal data should only be collected if necessary and relevant to an organization's data collection purpose.
Depicted by Table \ref{tab:requirements}, data minimization is our third operationalized principles.
As {\corp} collects a large assortment of data and data types, it is onerous for a developer to manually verify whether the organization is collecting more personal data than originally intended.

In contrast, we chose not to operationalize three principles (i.e. lawfulness, fairness, and transparency, accuracy, and purpose limitation), because these principles are more subjective in nature and/or have less applicability for {\corp}. 
For instance, the accuracy principle prescribes that personal data must be kept up to date and inaccurate personal data is fixed or erased \cite{gdpr_regulations}.
Personal data collected by {\corp} is pseudo-anonymous; if data was inaccurate for any particular reason, {\corp} has minimal ability to identify the corresponding data subject and the data subject is almost certainly not going to be affected.
Only {\corp}'s partners may be affected as they desire accurate data.

\subsubsection{Iterative Development and Evaluation of Requirements as Operationalized Requirements of GDPR Principles}
After determining three relevant principles, these principles were operationalized into privacy requirements as shown by Table \ref{tab:requirements}.
Specifically, the privacy requirements were developed based on input from {\corp} and implications of these principles on {\corp}'s various infrastructure resources. 
Hence, these requirements are relevant to {\corp}.
Moreover, we operationalized the integrity and confidentiality principle for {\corp}'s infrastructure including servers, load balancers, and databases.
Resulting requirements include ensuring that an access management resource does not provide a blanket policy that grants unrestricted access or action and a database is encrypted. 
Similarly, we applied the storage limitation and data minimization principles to databases, which are heavily used by {\corp} to store its data. 
Considering storage limitation, {\corp} automatically removes any long existing data.
Likewise, data minimization was refined into the privacy requirement, ``A database must not collect personal data types outside an organization's data collection purpose."

Our list of privacy requirements was iteratively refined based on the combined feedback from {\corp} and the results of operationalizing the requirements in a GDPR tool, which we discuss in more detail in section \ref{gdpr_tool}. 
We evaluated each requirements based on two properties: 1) a requirement is important to {\corp} 2) a requirement is derived from a GDPR principle. 
For example, {\corp} disagreed and felt the requirement ``Each load balancer must only use secure protocols" caused our GDPR tool to identify many load balancers that {\corp} perceived as otherwise secure. 
In response, we revised the NFR to ``Each load balancer must use secure protocols" to account for cases where a load balancer listened to both http and https traffic. 
The previous example represented a requirement that fulfilled the second property, but was not initially important enough to {\corp} as {\corp} felt the requirement was too stringent. 

On the contrary, an operationalized requirement is determined to be effective and valid when the requirement comes from a GDPR principle and {\corp} finds the requirement important.
For instance, an employee exclaimed, \q{It is peculiar that [redacted]...that should have all been fixed a while ago!}
Moreover, we also refined our privacy requirements based on lessons learned from external events. 
For example, when the Capital One breach occurred \cite{cloudsploit_technical_nodate}, that largely originated from misconfigurations of cloud infrastructure, we created requirements that applied to access and modification rights.

\subsection{Automated Testing of GDPR Requirements using a GDPR Tool} \label{gdpr_tool}
From the list of privacy requirement from Table \ref{tab:requirements}, we developed our second design science artifact: a GDPR tool that verifies these requirements and can be executed automatically on {\corp}'s system.
Specifically, our GDPR tool entails a series of Python scripts tailored for AWS.
In short, our tool checked privacy requirements in various elements on {\corp}'s AWS cloud infrastructure.
More importantly, our tool provided a vehicle for us to apply our operationalized requirements in practice and validate whether these requirements are reasonable and legitimate.

Assuming GDPR exposures in {\corp}'s infrastructure are found, our GDPR tool produces a list with detailed information about each exposure, such as location, name, ID, type of resource, and pertinent GDPR principle, which allows a developer to investigate the exposure in more detail.
Furthermore, our GDPR tool ran without requiring a human intervening to trigger an execution as Jenkins, a CI tool, triggers the GDPR tool to run weekly.
If {\corp} desired, {\corp} could run the tool every minute. 

\subsubsection{Iterative Development and Evaluation of GDPR Tool}
Our GDPR tool serves to realize our privacy NFRs in practice, which allows us to iteratively improve our privacy requirements and the tool itself. 
In the iterative development and evaluation of our GDPR tool, we received feedback from {\corp} in meeting, discussing, and analyzing the results produced by the tool.
The feedback helped evaluate the accuracy of the tool, as well as improve the efficiency of the  tool. 
For instance, when the GDPR found eleven load balancers of a specific type, we manually verified that there were eleven load balancers from the third party provider. 
We also modified our GDPR tool to reflect any changes to our list of privacy requirements.
Ultimately, the purpose of our evaluation was to ensure that our list of privacy requirements was verified in an automated tool.
The tool in turn automatically checked for potential GDPR exposures and provided meaningful details that can help an employee investigate an exposure.
Unfortunately, {\corp} did not consistently create tasks to address identified potential problems found by our tool during the course of study.
It may be that employees were currently limited by time and felt the potential GDPR exposures identified by the GDPR tool were not ``severe" enough to cause a drastic penalty if temporarily not investigated and resolved. 
However, we were encouraged by {\corp} agreement that our GDPR tool's results should have been added to the organization's backlog, but employees have been limited with other work. 

%% file: 08_discussion.tex
\section{Discussion and Implications}
Our tool and operationalized requirements have not addressed every GDPR challenge. We discuss ongoing challenges and what practitioner and research implications they bring.

\subsection{Time and motivation limit use of continuous GDPR compliance}
Continuous compliance \cite{fitzgerald_continuous_2014} is characterized as automatically checking regulatory compliance after each sprint.
Based on continuous compliance, if any non-compliance issues exist, the organization will add the list of issues to the organization's backlog to reduce the chances of the same non-compliance issue continuously recurring.
Next, the privacy tasks would be assigned a high priority and resolved in an subsequent sprint.
Since our GDPR tool executed on a continuous basis and was automated, we also had the opportunity to explore GDPR continuous compliance at {\corp}.

Our GDPR tool could serve as the first step of the continuous compliance process, whereby the tool executed at least once during each sprint and produced an actionable list of potential GDPR exposures. 
However, {\corp} did not regularly add tasks based on the created list of potential GDPR exposures. 
From our experience, there may be two interlinked causes to the inadequate adoption of continuous compliance in {\corp}.
First, employees are constantly busy and finding time to translate GDPR tool results into tasks and subsequently working on such tasks is overly time consuming.
This reason is supported by {\corp}'s continuous reassurance that our GDPR tool's results should have been added to the organization's backlog, but employees have been busy with other work.
Second, since time is such a valuable resource, employees may feel that the potential GDPR exposures identified by the GDPR tool are not ``severe" enough to cause a drastic penalty if temporarily not investigate and resolved. 
When employees have free time in the future, they could theoretically allocate time to treating and managing the results of our GDPR tool.
Moreover, the fact that GDPR continuous compliance has not yet been achieved in {\corp} does not mean that such feat is not possible in the future. 
As {\corp} hires more employees and matures, it is quite realistic that an employee may be tasked with mending the incomplete continuous compliance cycle and successfully conducting GDPR continuous compliance.

\subsubsection{Implications}
First, reliance on manual testing is major challenge to GDPR compliance, but operationalizing GDPR regulations with automated tools is a solution to alleviating manual testing.
Additional research into operationalizing other GDPR principles or rights may be a beneficial area of study.
\begin{myboxi}[Research Implication 1]
How to operationalize and automatically test compliance with the remaining GDPR regulations?
\end{myboxi}

Through operationalization of GDPR regulations, manual testing of GDPR compliance should be replaced with automated verification without the risk of human error or wasting time. This also ties in with organizational shifts to more continuous and automated software development \cite{fitzgerald_continuous_2017}.
However, given the difficulty in producing a tool that covers every aspect of privacy, an organization is best served by a complement of tools that together help support an organization's effective treatment of privacy. 
Continuous compliance is an excellent strategy for GDPR compliance, but buy-in from employees is required, and adhering to continuous compliance steps becomes habitual. 
CI can help an organization quickly respond to an issue, privacy included, but an organization may only realize these benefits if potential GDPR exposures are translated into work tasks and subsequently prioritized and resolved.  
\begin{myboxi}[Practitioner Implication 1]
Operationalize GDPR principles into relevant privacy requirements and use automated tests to continuously verify these requirements.
\end{myboxi}

\subsection{Insufficient knowledge management impedes privacy awareness and compliance}
Employees need to have a reasonable level of understanding of the GDPR in order to manage privacy requirements.
However, {\corp} did not have systematic employee training on the GDPR, nor explicit GDPR policies.
Consequently, an employee had to conduct individual research on the GDPR and knowledge was disseminated on an ad-hoc basis.
Furthermore, an employee who was not assigned work on privacy requirements was unlikely to have substantial awareness of the GDPR.
Given the lack of a standard of privacy knowledge that an employee must exhibit, an organization's overall ability to treat privacy was limited.
Raising privacy awareness in resource constrained organizations, especially in early development is difficult \cite{ataei_complying_2018}. Our observation shows that privacy awareness is not only challenging early on, but also difficult in the present and long term.
Privacy is an NFR that cannot be partially satisfied. If even one employee is not aware of the GDPR, they can easily cause privacy problems, even accidentally.
As demonstrated by cases like the Capital One data leak, one minor configuration mishap in a system's infrastructure can result in the exploitation of millions of users' sensitive data \cite{cloudsploit_technical_nodate}. 

Ensuring that an organization's privacy processes and policies are transparent for all employees is also vital to the organization's sufficient treatment of privacy.
A lack of shared understanding at {\corp} was exemplified where employees were not always aware of privacy processes upstream or downstream of their own work. 
For example, one employee who works in data processing automatically assumes that privacy is out of scope for his work because collected data are already GDPR compliant. 
The employee believes that the data were ``treated", which means the employee does not need to worry or consider the GDPR.
Yet, the employee was unsure about any specific privacy safeguards upstream in the process.
To effectively manage privacy, automatically assuming GDPR compliance seems inadequate.
When asked about the privacy safeguards that exist in upstream processes, the employee was unsure. 
If transparency of workflow and greater shared understanding existed, the aforementioned scenario would likely be avoided. 

As previously stated, GDPR compliance requires effort from everyone in an organization.
For long term compliance, not only is some level of training regarding privacy necessary, but also frequent reiteration of knowledge and awareness to ensure sufficient long term privacy awareness. 
Standardizing privacy training for employees will help an organization ensure that its employees have a relatively equal background of knowledge.
Furthermore, breaking down barriers and increasing shared understanding between employees is paramount to increasing transparency. 
An employee should be aware of privacy safeguards, not only holistically across an organization, but also in relation to one's own work.

\subsubsection{Implications}
Insufficient awareness and knowledge management may impede long term compliance because a developer is less likely to be able to address a potential privacy problem if the developer is unaware of the pertaining privacy regulation.
Therefore, more research is needed to find more efficient strategies to disseminate the GDPR's implications to improve an organization's ability to handle compliance.
\begin{myboxi}[Research Implication 2]
How to efficiently and effectively disseminate and manage GDPR knowledge? How can existing GDPR knowledge be updated with new best practices and patterns?
\end{myboxi}

For effective GDPR compliance, an organization needs to ensure that each employee is adequately trained and aware of the GDPR.
An employee must be cognizant of a regulation to consider the regulation during work. 
Processes and privacy safeguards also must be transparent so that an employee has a breadth of knowledge of the safeguards across an organization.
\begin{myboxi}[Practitioner Implication 2]
Each employee must be adequately trained about the GDPR. Organizations should develop GDPR policies specifically relevant to developers. 
\end{myboxi}

\subsection{Managers and developers have sharply different priorities for GDPR compliance}
{\corp} had a relatively sizeable difference in mentality with respect to the importance and relevance of privacy between managers and developers.
In particular, managers value privacy more than developers.
This observation was shown by our first survey where managers ranked privacy as the most important NFR to {\corp}'s business whereas developers ranked privacy sixth.
A second round of surveys occurred seven months after the first survey.
In subsequent surveys, managers continued to rank privacy as the most important NFR for {\corp}'s business, but developer responses were little changed. 
This is in spite of increased awareness by all employees due an external GDPR compliance audit performed by consultants at the time, as well as our ongoing study.

The unequal distribution of privacy tasks between employees may have contributed to the inequality in valuing privacy.
In general, managers were more associated with privacy than developers. 
While privacy may be a ``nice to have" quality of a feature, privacy does not directly affect developer work as much as as an NFR like reliability. 
A developer is unlikely to release a feature if the new feature degrades the reliability of the overall software.
Privacy compliance is unlikely to immediately affect the development and release of a feature.
Hence, developers may feel less connection to privacy, which may hurt the long term commitment to compliance. 

The imbalance in valuing privacy between developers and management reduces the quality of an organization's GDPR compliance, especially long term compliance.
As time passes, fewer employees become involved with privacy work, which entails lower awareness and familiarity of privacy.
For the purposes of effectively treating privacy using strategies including privacy by design (PbD) \cite{cavoukian2010privacy}, developers play a pivotal role \cite{hadar_privacy_2018}.
While consultants and lawyers may help interpret and provide guidance on GDPR regulations, developers are ultimately assigned to convert these regulations into requirements and realize the requirements in software.
Without developer involvement, an organization's compliance effort is futile.
Developers need to have a strong perception and understanding of privacy, but developers often relegate privacy measures to policy based solutions as opposed to architectural changes \cite{hadar_privacy_2018}.
Furthermore, organizational discouragement was previously found to be a significant barrier to motivating developers towards privacy \cite{hadar_privacy_2018}.
In contrast, {\corp's} managers support adoption of privacy initiatives and prioritize privacy.
Ultimately, if managers are highly motivated and intend to promote GDPR compliance within an organization, this motivation is to no avail if people who help implement compliance do not share the same inspiration.
To advance towards long term GDPR compliance, an emphasis on privacy must be shown by both managers and developers (i.e. from the ``grassroots" level). 
P1 (manager) stated, \q{For the entire organization to be GDPR compliant, each individual has to be GDPR compliant and review needs to be done on an individual basis rather than just an organizational perspective}.  

\subsubsection{Implications}
The difference in mentality between developers and managers towards privacy may be a long term challenge to compliance. 
Breaking down the barriers between developers and managers and increasing the shared understanding of privacy may be a judicious research exploration.
\begin{myboxi}[Research Implication 3]
How to break down barriers and increase shared understanding between developers and managers?
\end{myboxi}

To ensure that compliance work is effectively implemented and employees carefully treat privacy, the silo between manager and developer roles must be reduced. 
As developers ultimately carry out many privacy tasks and interact with data on an everyday basis, a developer must share the same level of urgency and value towards privacy.
\begin{myboxi}[Practitioner Implication 3]
Continuous compliance with GDPR (e.g. via automated testing) may help to make privacy more of a day to day concern for developers.
\end{myboxi}

\subsection{Offloading privacy concerns relinquishes compliance control to others}
{\corp} heavily relies on third party services such as AWS, Azure, and Google Cloud Platform. {\corp} believes these services provide privacy protection.
First, services, like AWS, provide state of the art cloud infrastructure, which naturally should have excellent GDPR compliance safeguards that would help protect {\corp}'s data.
Second, these services are in a unique position where they act as the de facto ``processors" of {\corp}'s data. 
Per the GDPR regulations, these processors share the same responsibilities as {\corp} to store and process the data in a GDPR compliant manner. 
Any potential criticism of {\corp} would almost certainly apply to a corresponding third party service as well. 
By using services provided by large organizations, {\corp} has effectively offloaded or otherwise shared large aspects of privacy responsibility with these third parties.

However, even industry leading third party services may have vulnerabilities. 
Third party services may not be fully GDPR compliant and organizations typically still need to self manage the configuration of its cloud infrastructure.


Throughout {\corp}'s compliance process, {\corp} hired reputable privacy consultants to review compliance and suggest improvements.
As described by P4, ``[Reputable consultant] reviewed our compliance and he was impressed'' (and hence, P4 saw no need to do anything else).
Solely relying on the positive review of a consultant may provide a false sense of assurance and decrease motivation to further improve the organization's compliance. 
In addition, consultants often provide differing opinions on the same issue or miss aspects in a review \cite{elga2007reflection}. 
The compliance certificate provided to {\corp} is perceived as reputable, but compliance review may be subjective as there is not a universal standard method or framework to conduct a GDPR compliance audit. Ultimately true compliance is determined in the breach, i.e., when faced with regulatory or legal action.

Yet, over reliance on these partners to take the brunt of any potential scrutiny may over expose {\corp} to external forces. 
Instead of having full control over its GDPR compliance, the organization may be unintentionally exposing itself to the compliance of its third party providers and trusting that partners will support {\corp}. 

\subsubsection{Implications}
Partnering with large entities and working with reputable consultants may facilitate offloading or at least sharing privacy responsibilities with these external entities.
However, offloading privacy responsibility may result in losing control to external entities. 
Research is needed into the risks involved with offloading privacy responsibilities and also managing these risks. 
\begin{myboxi}[Research Implication 4]
How can we quantify the risks of offloading privacy responsibilities? Is there a repeatable way to make decisions based on this analysis?
\end{myboxi}

An organization can seemingly offload aspects of privacy responsibility to external entities. 
For example, when an organization uses cloud infrastructure from a third party service, the organization should inherently receive some privacy protections for its data. 
However, offloading may be a cause for concern because the organization must review whether or not the third party service is sufficiently treating privacy. 
\begin{myboxi}[Practitioner Implication 4]
An organization may offload privacy responsibilities to external entities, but the organization must be cognizant that offloading has risks. 
\end{myboxi}




%% file: 09_related_work.tex
\section{Background and Related Work}
\subsection{Privacy Regulations}
As the replacement to the 1995 EU Data Protection Directive, the GDPR provided organizations a two year grace period to comply \cite{gdpr_general_site}.
Since the GDPR united privacy regulations in the EU under one umbrella regulation, the GDPR removed the need to tweak treatment of privacy for each EU country.

In addition to the the six main GDPR data processing principles and the accountability principle mentioned earlier in this paper, an organization must also consider a plethora of rights given to each data subject, such as right to be informed and right to object \cite{ico}. 
As part of compliance, an organization must adhere to each principle and right. 
After all, a data subject may report an organization to a DPA for a single instance of non-compliance whether the instance is an infraction on a right or principle. 
As a trailblazing privacy law, the GDPR has influenced other governments to pass similar laws.
In particular, some US states have passed laws such as the CCPA\footnote{\url{https://leginfo.legislature.ca.gov/faces/billTextClient.xhtml?bill_id=201720180AB375}}and SHIELD Act\footnote{\url{https://www.nysenate.gov/legislation/bills/2019/s5575}}. 
Specifically, the CCPA mimics and strives to be similarly encompassing as the GDPR.
However, unlike the GDPR, the CCPA has no upper limit on the amount that an organization can be fined; an organization may be fined for up to billions of dollars. 
The recent enactment of laws may represent an increase in privacy laws in the future; complying with the GDPR should help prepare an organization with future privacy regulations as well.

\subsection{Privacy Tools and Methodologies}
Over the past few decades, the increased integration of privacy and technology has produced privacy enhancing technologies (PETs) and privacy by design (PbD), which aim to increase privacy in software \cite{cavoukian2010privacy}.
PETs strive to use technology to protect the privacy of individual or groups of individuals \cite{heurix2015taxonomy}.
Some PETs include protecting user identities \cite{pfitzmann_anonymity_nodate} and anonymizing network data \cite{Dingledine:2004:TSO:1251375.1251396}.
In contrast, PbD not only calls for prioritization of privacy from the onset of an organization \cite{cavoukian2010privacy}, but also during planning, operations, and development phases of a software's life cycle.
However, the extent of PbD's prioritization may be dampened due to situations where ``developers are actively discouraged from making informational privacy a priority" \cite{hadar_privacy_2018}. 
To strive towards optimal treatment of privacy, organizations should include positive reinforcements and motivate developers to increase their value of privacy \cite{hadar_privacy_2018}.

Other privacy methodologies include Deng et al's \cite{deng_privacy_2011} LINDDUN Methodology, which aims to identify privacy threats in a system through analysis of the system's data flow diagram.
However, analysing data flow diagrams means that LINDDUN primarily provides a high level analysis of privacy threats as opposed to specific implementation details \cite{deng_privacy_2011}.
Yet, one solution strategy discussed by Deng et al. \cite{deng_privacy_2011}, of removing or shutting off system elements to decrease risk was observed in our work;
{\corp} shut down potentially concerning elements of its system before the GDPR deadline to decrease risk and hassle.
When risk of an element is excessively difficult to mitigate, removing the element is the safest approach. 

Like PbD, other privacy methodologies exist to enhance privacy in an engineering context, but organizational commitment from the inception of a system is needed as delayed focus on privacy may be too late \cite{gurses2011engineering}.
Privacy-by-policy and privacy-by-architecture are two approaches \cite{privacypractices} to protect privacy.
Privacy-by-policy is the concept of modifying a system to suit privacy, often using privacy policies and user choice as mechanisms. 
Privacy-by-architecture is the notion to fundamentally incorporate privacy into a system \cite{privacypractices}; user data are anonymous and efforts to exploit user data is futile \cite{privacypractices}.
Privacy-by-policy is less reliable and robust, but is frequently adopted by businesses due to its convenience, as well as being a popular choice among developers \cite{hadar_privacy_2018}. 
Privacy-by-architecture is more reliable, but it has stringent privacy expectations and may not be easily adaptable to a pre-existing system \cite{privacypractices}. 

\subsection{Current GDPR Challenges and State of Research}
One year post GDPR deadline, many organizations are still not GDPR compliant \cite{narendra_almost_2019} and/or may never be fully compliant \cite{chantzos_gdpr_2019}.
In particular, smaller organizations that did not previously take appropriate security and privacy measures, like PbD, may feel burdened by GDPR compliance \cite{sirur_are_2018}. 

Multiple frameworks have been suggested to assist GDPR compliance.
Brodin proposes a framework with steps to guide an organization to compliance \cite{brodin_framework_2019}, but the framework is relatively high level and lacks details such as how an organization may implement each step.
Similarly, a six step approach was proposed to help an organization elicit solution requirements from the GDPR \cite{ayala-rivera_grace_2018}; the appropriateness of the requirements were validated with privacy experts, but the requirements lacked clear cut measurables for validation. 
In contrast, our requirements are more discernable, which allowed us to operationalize requirements into an automated tool.

Coles et al. \cite{coles_tool-supporting_2018} described a tool supported approach to performing a data protection impact assessment (DPIA), which is one method to demonstrate compliance. 
Our research focused on a different aspect of compliance, which is helping to achieve compliance as opposed to demonstrating compliance was achieved. 
Holistically analyzing the GDPR, Tikkinen-Piri et al. \cite{tikkinen-piri_eu_2018} found twelve ramifications that an organization must be cognizant and called for more empirical research into GDPR compliance and challenges.
Our study answers this call for further research into GDPR compliance practices and challenges; we found three challenges to compliance and four hindrances to long term compliance.

After interviewing six experts involved with implementing the GDPR, Ataei et al. \cite{ataei_complying_2018} found three compliance challenges related to user interfaces of location based services.
One of the challenges was also awareness, but their focus was on raising awareness early in development.
In contrast, the challenge of awareness in our research refers to awareness throughout the life cycle of an organization and software, not just in early stage development. 

Regarding user rights, Altorbaq et al. \cite{altorbaq_data_2017} conducted ten interviews to formulate guidance for adherence to GDPR data subject rights, which include twelve challenges and fourteen recommendations grouped by stages of a personal information life cycle model created by the same authors.
For service oriented SMEs, a study mapped a set of requirements generated from constraints that applied to the studied organization as a result of the GDPR and modified the SME's architecture to satisfy these requirements \cite{hjerppe_general_2019}. 
In contrast, our work focuses on the efficacy of ``operationalizing" privacy requirements derived from a set of GDPR principles, in an automated tool, and in a CI context.


One quality organizations and researchers often vent frustration is the ambiguity of the GDPR, but Cool \cite{cool_impossible_2019} explained that the GDPR is intentionally vague because it must anticipate future technologies. 
Nonetheless, this ambiguity creates difficulties for organizations as observed in our study.
Ringmann et al. \cite{ringmann_requirements_2018} defined technical requirements that served to help make a software compliant, but the requirements are relatively generic as the authors wanted the requirements to apply to as many organizations as possible \cite{ringmann_requirements_2018}.
Static code analysis is suggested as a method to identify potential GDPR exposures, but static analysis is limited to code not other candidates for non-compliance like infrastructure or policies \cite{Ferrara2018StaticAF}.
Continuous compliance prescribes continuous verification of a software for regulatory compliance. 
Satisfying the continuous verification may render the necessity for applying a multitude of automated security and privacy tools, like our GDPR tool that raise awareness about possible GDPR exposures.
Other available tool include
IBM's Guardium Analyzer \cite{Shah:2018:ISG:3291291.3291349} and HPE's GDPR Starter Kit\footnote{https://www.hpe.com/us/en/newsroom/blog-post/2018/06/get-ready-for-the-gdpr-with-hpe.html}.
Regardless, as we observed in our study, flagging potential non-compliant candidates is only one step towards continuous compliance.

%% file: 11_threats_to_validity.tex
\section{Limitations}
The study has a few primary limitations.

To ensure credibility of the report, we make our interview guide available, and use thick descriptions of the interview approaches. However, our confidentiality agreement with our partner limits our ability to be completely transparent. At {\corp} we interviewed participants in a variety of roles to ensure we had a valid sample. The interview data we collected from our participants may be limited due to participants answering a question a certain way because they know they are being watched (the observer effect). A similar bias may apply to observational data we collected. We removed any themes that did have corroborating support from multiple interviewees. As part of our iterative, design science approach, we validated each challenge and operationalized GDPR requirement with employees at \corp{}. We triangulated our findings from interviews with our observational data, surveys, and limited analysis of code and issue tracker artifacts.

The interpretation of research results may be subject to researcher bias as one co-author has extensive knowledge about {\corp}. To some extent this is inherent in the ethnographic method. In our view, the extensive knowledge merely served to provide context about {\corp}, not to bias any inferences or conclusions of results.

We secured institutional ethical review approval prior to our study. We also reminded participants that we were not at the company to judge or find blame, that they would be anonymous, and that our research goal was focused on the company and GDPR, not individuals.

Clearly the usefulness and generalizability of the paper may be limited as we studied a single company. 
The single company is a small organization (several dozen employees) and operates in a data gathering business, with a reliance on cloud infrastructure. 
However, although we may not be able to generalize our study to other settings, such as a large organization developing safety critical software, we expect organizations of similar size and context (e.g. GDPR-applicable, cloud-based, CI-practicing) to encounter similar challenges as our collaborator. While we identified and helped resolve specific problems for one organization, this focus allowed us to develop deep insight into the company and multiple iterations on the tool, increasing credibility of our results.

%% file: 12_conclusions.tex
\section{Conclusion}
The practices and challenges of GDPR compliance in small organizations practicing CI is still a relatively unexplored area of research. 
Through design science research, we investigated the compliance challenges in a small organization and identified three primary challenges. 
We focused on alleviating the challenge of relying on manual GDPR tests and produced two design science artifacts: operationalized GDPR requirements and an automated GDPR tool. 

We derived four implications for research and four implications for practice.
Researchers may consider further operationalizing and automatically testing compliance of other GDPR regulations. 
Researchers may also investigate quantifying the risks of offloading privacy responsibilities and finding a repeatable way to make decisions based on this analysis. 
Similarly, an organization can operationalize GDPR principles into requirements and continuously test these requirements. 
Finally, an organization can offload privacy to third parties with the caveat that the organization may lose some control of privacy.

%% file: main.bbl
\begin{thebibliography}{10}
\providecommand{\url}[1]{#1}
\csname url@samestyle\endcsname
\providecommand{\newblock}{\relax}
\providecommand{\bibinfo}[2]{#2}
\providecommand{\BIBentrySTDinterwordspacing}{\spaceskip=0pt\relax}
\providecommand{\BIBentryALTinterwordstretchfactor}{4}
\providecommand{\BIBentryALTinterwordspacing}{\spaceskip=\fontdimen2\font plus
\BIBentryALTinterwordstretchfactor\fontdimen3\font minus
  \fontdimen4\font\relax}
\providecommand{\BIBforeignlanguage}[2]{{%
\expandafter\ifx\csname l@#1\endcsname\relax
\typeout{** WARNING: IEEEtran.bst: No hyphenation pattern has been}%
\typeout{** loaded for the language `#1'. Using the pattern for}%
\typeout{** the default language instead.}%
\else
\language=\csname l@#1\endcsname
\fi
#2}}
\providecommand{\BIBdecl}{\relax}
\BIBdecl

\bibitem{gdpr_regulations}
\BIBentryALTinterwordspacing
``\BIBforeignlanguage{en}{Regulation ({EU}) 2016/679 of the {European}
  {Parliament} and of the {Council} of 27 {April} 2016 on the protection of
  natural persons with regard to the processing of personal data and on the
  free movement of such data, and repealing {Directive} 95/46/{EC} ({General}
  {Data} {Protection} {Regulation}) ({Text} with {EEA} relevance)},''
  \url{http://data.europa.eu/eli/reg/2016/679/oj/eng}, May 2016. [Online].
  Available: \url{http://data.europa.eu/eli/reg/2016/679/oj/eng}
\BIBentrySTDinterwordspacing

\bibitem{npr_article}
\BIBentryALTinterwordspacing
D.~Rijo. (2019, May) \BIBforeignlanguage{en-US}{{NPR} \& {GDPR}: {Users} that
  decline cookies sent to a plain text website}.
  \url{https://ppc.land/npr-gdpr-users-that-decline-cookies-sent-to-a-plain-text-website/}.
  [Online]. Available:
  \url{https://ppc.land/npr-gdpr-users-that-decline-cookies-sent-to-a-plain-text-website/}
\BIBentrySTDinterwordspacing

\bibitem{nbc_block}
\BIBentryALTinterwordspacing
(2019) E.{U}. privacy regulations cause some web services to block {European}
  visitors. [Online]. Available:
  \url{https://www.nbcnews.com/tech/tech-news/chicago-tribune-los-angeles-times-block-european-users-due-gdpr-n877591}
\BIBentrySTDinterwordspacing

\bibitem{aranda_requirements_2007}
\BIBentryALTinterwordspacing
J.~Aranda, S.~Easterbrook, and G.~Wilson,
  ``\BIBforeignlanguage{en}{Requirements in the wild: {How} small companies do
  it},'' in \emph{\BIBforeignlanguage{en}{15th {IEEE} {International}
  {Requirements} {Engineering} {Conference} ({RE} 2007)}}.\hskip 1em plus 0.5em
  minus 0.4em\relax Delhi, India: IEEE, Oct. 2007, pp. 39--48. [Online].
  Available: \url{http://ieeexplore.ieee.org/document/4384165/}
\BIBentrySTDinterwordspacing

\bibitem{ries_lean_2011}
E.~Ries, \emph{\BIBforeignlanguage{English}{The {Lean} {Startup}: {How}
  {Today}'s {Entrepreneurs} {Use} {Continuous} {Innovation} to {Create}
  {Radically} {Successful} {Businesses}}}, 1st~ed.\hskip 1em plus 0.5em minus
  0.4em\relax New York: Currency, Sep. 2011.

\bibitem{gralha_evolution_2018}
C.~Gralha, D.~Damian, A.~I.~T. Wasserman, M.~Goulão, and J.~Araújo, ``The
  {Evolution} of {Requirements} {Practices} in {Software} {Startups},'' in
  \emph{Proceedings of the 40th {International} {Conference} on {Software}
  {Engineering}}, 2018, pp. 823--833.

\bibitem{glinz}
M.~{Glinz}, ``On non-functional requirements,'' in \emph{International
  Conference on Requirements Engineering}, Oct 2007, pp. 21--26.

\bibitem{ramesh_agile_2010}
B.~Ramesh, L.~Cao, and R.~Baskerville, ``Agile requirements engineering
  practices and challenges: an empirical study,'' \emph{Information Systems
  Journal}, vol.~20, no.~5, pp. 449--480, 2010.

\bibitem{narendra_almost_2019}
\BIBentryALTinterwordspacing
M.~Narendra. (2019, Jul.) \BIBforeignlanguage{en-US}{Almost a third of {EU}
  firms still not {GDPR} compliant}. [Online]. Available:
  \url{https://gdpr.report/news/2019/07/22/almost-a-third-of-eu-firms-still-not-gdpr-compliant/}
\BIBentrySTDinterwordspacing

\bibitem{chantzos_gdpr_2019}
\BIBentryALTinterwordspacing
I.~Chantzos. (2019, May) {GDPR} {Turns} 1: {Many} {Companies} {Still} {Not}
  {Ready}. [Online]. Available:
  \url{https://www.symantec.com/blogs/expert-perspectives/gdpr-turns-1-many-companies-still-not-ready}
\BIBentrySTDinterwordspacing

\bibitem{sirur_are_2018}
\BIBentryALTinterwordspacing
S.~Sirur, J.~R.~C. Nurse, and H.~Webb, ``Are we there yet? {Understanding} the
  challenges faced in complying with the {General} {Data} {Protection}
  {Regulation} ({GDPR}),'' \emph{arXiv:1808.07338 [cs]}, Aug. 2018, arXiv:
  1808.07338. [Online]. Available: \url{http://arxiv.org/abs/1808.07338}
\BIBentrySTDinterwordspacing

\bibitem{poritskiy_benefits_2019}
N.~Poritskiy, F.~Oliveira, and F.~Almeida, ``The benefits and challenges of
  general data protection regulation for the information technology sector,''
  \emph{Digital Policy, Regulation and Governance}, vol. ahead-of-print, Sep.
  2019.

\bibitem{hevner_design_2004}
A.~R. Hevner, S.~T. March, J.~Park, and S.~Ram,
  ``\BIBforeignlanguage{en}{Design {Science} in {Information} {Systems}
  {Research}},'' \emph{\BIBforeignlanguage{en}{Management Information Systems
  Quarterly}}, p.~32, Mar. 2004.

\bibitem{eu_sme}
\BIBentryALTinterwordspacing
(2019) What is an {SME}? {\textbar} {Internal} {Market}, {Industry},
  {Entrepreneurship} and {SMEs}. [Online]. Available:
  \url{https://ec.europa.eu/growth/smes/business-friendly-environment/sme-definition_en}
\BIBentrySTDinterwordspacing

\bibitem{sedlmair_design_2012}
\BIBentryALTinterwordspacing
M.~Sedlmair, M.~Meyer, and T.~Munzner, ``\BIBforeignlanguage{en}{Design {Study}
  {Methodology}: {Reflections} from the {Trenches} and the {Stacks}},''
  \emph{\BIBforeignlanguage{en}{IEEE Transactions on Visualization and Computer
  Graphics}}, vol.~18, no.~12, pp. 2431--2440, Dec. 2012. [Online]. Available:
  \url{http://ieeexplore.ieee.org/document/6327248/}
\BIBentrySTDinterwordspacing

\bibitem{Dustin:1999:AST:310674}
E.~Dustin, J.~Rashka, and J.~Paul, \emph{Automated Software Testing:
  Introduction, Management, and Performance}.\hskip 1em plus 0.5em minus
  0.4em\relax Boston, MA, USA: Addison-Wesley Longman Publishing Co., Inc.,
  1999.

\bibitem{cool_impossible_2019}
\BIBentryALTinterwordspacing
A.~Cool, ``\BIBforeignlanguage{en}{Impossible, unknowable, accountable:
  {Dramas} and dilemmas of data law},'' \emph{\BIBforeignlanguage{en}{Social
  Studies of Science}}, vol.~49, no.~4, pp. 503--530, Aug. 2019. [Online].
  Available: \url{https://doi.org/10.1177/0306312719846557}
\BIBentrySTDinterwordspacing

\bibitem{cloudsploit_technical_nodate}
\BIBentryALTinterwordspacing
{CloudSploit}, ``A {Technical} {Analysis} of the {Capital} {One} {Hack} -
  {CloudSploit}.'' [Online]. Available:
  \url{https://blog.cloudsploit.com/a-technical-analysis-of-the-capital-one-hack-a9b43d7c8aea}
\BIBentrySTDinterwordspacing

\bibitem{fitzgerald_continuous_2014}
\BIBentryALTinterwordspacing
B.~Fitzgerald and K.-J. Stol, ``Continuous {Software} {Engineering} and
  {Beyond}: {Trends} and {Challenges},'' in \emph{Proceedings of the 1st
  {International} {Workshop} on {Rapid} {Continuous} {Software} {Engineering}},
  ser. {RCoSE} 2014.\hskip 1em plus 0.5em minus 0.4em\relax New York, NY, USA:
  ACM, 2014, pp. 1--9, event-place: Hyderabad, India. [Online]. Available:
  \url{http://doi.acm.org/10.1145/2593812.2593813}
\BIBentrySTDinterwordspacing

\bibitem{fitzgerald_continuous_2017}
\BIBentryALTinterwordspacing
------, ``Continuous software engineering: {A} roadmap and agenda,''
  \emph{Journal of Systems and Software}, vol. 123, pp. 176--189, Jan. 2017.
  [Online]. Available:
  \url{http://www.sciencedirect.com/science/article/pii/S0164121215001430}
\BIBentrySTDinterwordspacing

\bibitem{ataei_complying_2018}
\BIBentryALTinterwordspacing
M.~Ataei, A.~Degbelo, C.~Kray, and V.~Santos,
  ``\BIBforeignlanguage{en}{Complying with {Privacy} {Legislation}: {From}
  {Legal} {Text} to {Implementation} of {Privacy}-{Aware} {Location}-{Based}
  {Services}},'' \emph{\BIBforeignlanguage{en}{ISPRS International Journal of
  Geo-Information}}, vol.~7, no.~11, p. 442, Nov. 2018. [Online]. Available:
  \url{https://www.mdpi.com/2220-9964/7/11/442}
\BIBentrySTDinterwordspacing

\bibitem{cavoukian2010privacy}
A.~Cavoukian, ``Privacy by design: the definitive workshop,'' \emph{Identity in
  the Information Society}, vol.~3, no.~2, pp. 247--251, 2010.

\bibitem{hadar_privacy_2018}
\BIBentryALTinterwordspacing
I.~Hadar, T.~Hasson, O.~Ayalon, E.~Toch, M.~Birnhack, S.~Sherman, and
  A.~Balissa, ``\BIBforeignlanguage{en}{Privacy by designers: software
  developers’ privacy mindset},'' \emph{\BIBforeignlanguage{en}{Empirical
  Software Engineering}}, vol.~23, no.~1, pp. 259--289, Feb. 2018. [Online].
  Available: \url{http://link.springer.com/10.1007/s10664-017-9517-1}
\BIBentrySTDinterwordspacing

\bibitem{elga2007reflection}
A.~Elga, ``Reflection and disagreement,'' \emph{No{\^u}s}, vol.~41, no.~3, pp.
  478--502, 2007.

\bibitem{gdpr_general_site}
\BIBentryALTinterwordspacing
``\BIBforeignlanguage{en}{Data protection in the {EU}},''
  \url{https://ec.europa.eu/info/law/law-topic/data-protection/data-protection-eu_en}.
  [Online]. Available:
  \url{https://ec.europa.eu/info/law/law-topic/data-protection/data-protection-eu_en}
\BIBentrySTDinterwordspacing

\bibitem{ico}
\BIBentryALTinterwordspacing
``The principles.'' [Online]. Available:
  \url{https://ico.org.uk/for-organisations/guide-to-data-protection/guide-to-the-general-data-protection-regulation-gdpr/principles/}
\BIBentrySTDinterwordspacing

\bibitem{heurix2015taxonomy}
J.~Heurix, P.~Zimmermann, T.~Neubauer, and S.~Fenz, ``A taxonomy for privacy
  enhancing technologies,'' \emph{Computers \& Security}, vol.~53, pp. 1--17,
  2015.

\bibitem{pfitzmann_anonymity_nodate}
\BIBentryALTinterwordspacing
A.~Pfitzmann and M.~Hansen, ``\BIBforeignlanguage{en}{Anonymity,
  {Unlinkability}, {Undetectability}, {Unobservability}, {Pseudonymity}, and
  {Identity} {Management} – {A} {Consolidated} {Proposal} for
  {Terminology}},'' p.~83, 2008. [Online]. Available:
  \url{http://dud.inf.tu-dresden.de/Anon_Terminology.shtml}
\BIBentrySTDinterwordspacing

\bibitem{Dingledine:2004:TSO:1251375.1251396}
\BIBentryALTinterwordspacing
R.~Dingledine, N.~Mathewson, and P.~Syverson, ``Tor: The second-generation
  onion router,'' in \emph{Proceedings of the 13th Conference on USENIX
  Security Symposium - Volume 13}, ser. SSYM'04.\hskip 1em plus 0.5em minus
  0.4em\relax Berkeley, CA, USA: USENIX Association, 2004, pp. 21--21.
  [Online]. Available: \url{http://dl.acm.org/citation.cfm?id=1251375.1251396}
\BIBentrySTDinterwordspacing

\bibitem{deng_privacy_2011}
\BIBentryALTinterwordspacing
M.~Deng, K.~Wuyts, R.~Scandariato, B.~Preneel, and W.~Joosen,
  ``\BIBforeignlanguage{en}{A privacy threat analysis framework: supporting the
  elicitation and fulfillment of privacy requirements},''
  \emph{\BIBforeignlanguage{en}{Requirements Engineering}}, vol.~16, no.~1, pp.
  3--32, Mar. 2011. [Online]. Available:
  \url{https://doi.org/10.1007/s00766-010-0115-7}
\BIBentrySTDinterwordspacing

\bibitem{gurses2011engineering}
S.~G{\"u}rses, C.~Troncoso, and C.~Diaz, ``Engineering privacy by design,''
  \emph{Computers, Privacy \& Data Protection}, vol.~14, no.~3, p.~25, 2011.

\bibitem{privacypractices}
S.~{Spiekermann} and L.~F. {Cranor}, ``Engineering privacy,'' \emph{IEEE
  Transactions on Software Engineering}, vol.~35, no.~1, pp. 67--82, Jan 2009.

\bibitem{brodin_framework_2019}
\BIBentryALTinterwordspacing
M.~Brodin, ``\BIBforeignlanguage{en}{A {Framework} for {GDPR} {Compliance} for
  {Small}- and {Medium}-{Sized} {Enterprises}},''
  \emph{\BIBforeignlanguage{en}{European Journal for Security Research}},
  vol.~4, no.~2, pp. 243--264, Oct. 2019. [Online]. Available:
  \url{http://link.springer.com/10.1007/s41125-019-00042-z}
\BIBentrySTDinterwordspacing

\bibitem{ayala-rivera_grace_2018}
V.~Ayala-Rivera and L.~Pasquale, ``The {Grace} {Period} {Has} {Ended}: {An}
  {Approach} to {Operationalize} {GDPR} {Requirements},'' in \emph{2018 {IEEE}
  26th {International} {Requirements} {Engineering} {Conference} ({RE})}, Aug.
  2018, pp. 136--146, iSSN: 2332-6441, 1090-705X.

\bibitem{coles_tool-supporting_2018}
J.~Coles, S.~Faily, and D.~Ki-Aries, ``Tool-{Supporting} {Data} {Protection}
  {Impact} {Assessments} with {CAIRIS},'' in \emph{2018 {IEEE} 5th
  {International} {Workshop} on {Evolving} {Security} {Privacy} {Requirements}
  {Engineering} ({ESPRE})}, Aug. 2018, pp. 21--27.

\bibitem{tikkinen-piri_eu_2018}
\BIBentryALTinterwordspacing
C.~Tikkinen-Piri, A.~Rohunen, and J.~Markkula, ``\BIBforeignlanguage{en}{{EU}
  {General} {Data} {Protection} {Regulation}: {Changes} and implications for
  personal data collecting companies},'' \emph{\BIBforeignlanguage{en}{Computer
  Law \& Security Review}}, vol.~34, no.~1, pp. 134--153, Feb. 2018. [Online].
  Available:
  \url{http://www.sciencedirect.com/science/article/pii/S0267364917301966}
\BIBentrySTDinterwordspacing

\bibitem{altorbaq_data_2017}
A.~Altorbaq, F.~Blix, and S.~Sörman, ``Data subject rights in the cloud: {A}
  grounded study on data protection assurance in the light of {GDPR},'' in
  \emph{2017 12th {International} {Conference} for {Internet} {Technology} and
  {Secured} {Transactions} ({ICITST})}, Dec. 2017, pp. 305--310.

\bibitem{hjerppe_general_2019}
\BIBentryALTinterwordspacing
K.~Hjerppe, J.~Ruohonen, and V.~Leppänen, ``The {General} {Data} {Protection}
  {Regulation}: {Requirements}, {Architectures}, and {Constraints},''
  \emph{arXiv:1907.07498 [cs]}, Jul. 2019, arXiv: 1907.07498. [Online].
  Available: \url{http://arxiv.org/abs/1907.07498}
\BIBentrySTDinterwordspacing

\bibitem{ringmann_requirements_2018}
S.~D. Ringmann, H.~Langweg, and M.~Waldvogel,
  ``\BIBforeignlanguage{en}{Requirements for {Legally} {Compliant} {Software}
  {Based} on the {GDPR}},'' in \emph{\BIBforeignlanguage{en}{On the {Move} to
  {Meaningful} {Internet} {Systems}. {OTM} 2018 {Conferences}}}, ser. Lecture
  {Notes} in {Computer} {Science}, H.~Panetto, C.~Debruyne, H.~A. Proper, C.~A.
  Ardagna, D.~Roman, and R.~Meersman, Eds.\hskip 1em plus 0.5em minus
  0.4em\relax Springer International Publishing, 2018, pp. 258--276.

\bibitem{Ferrara2018StaticAF}
P.~Ferrara and F.~Spoto, ``Static analysis for {GDPR} compliance,'' in
  \emph{Italian Conference on Cyber Security}, 2018.

\bibitem{Shah:2018:ISG:3291291.3291349}
\BIBentryALTinterwordspacing
D.~Shah, L.~Lindsay, J.~Diaz, S.~Shechter, and A.~Becher, ``Ibm security
  guardium analyzer bootcamp,'' in \emph{Proceedings of the 28th Annual
  International Conference on Computer Science and Software Engineering}, ser.
  CASCON '18.\hskip 1em plus 0.5em minus 0.4em\relax Riverton, NJ, USA: IBM
  Corp., 2018, pp. 380--382. [Online]. Available:
  \url{http://dl.acm.org.ezproxy.library.uvic.ca/citation.cfm?id=3291291.3291349}
\BIBentrySTDinterwordspacing

\end{thebibliography}
